\documentclass{article}

\usepackage{arxiv}

\usepackage{hyperref}       
\usepackage{url}            
\usepackage{booktabs}       
\usepackage{amsfonts}       
\usepackage{nicefrac}       
\usepackage{microtype}      
\usepackage[utf8]{inputenc}
\usepackage[english]{babel}
\usepackage{amsmath,amsfonts,amssymb}
\usepackage{lmodern}
\usepackage[scaled]{helvet}

\usepackage{apacite}

\usepackage[T1]{fontenc}
\usepackage{lettrine} 

\usepackage{graphicx}

\title{The Cultural Transmission of Tacit Knowledge}

\author{
  Helena Miton\\
Santa Fe Institute \\
1399 Hyde Park Road \\
Santa Fe, NM 87501 USA \\
   \And
 Simon DeDeo \\
Social \& Decision Sciences \\
Carnegie Mellon University \\ 5000 Forbes Avenue \\
Pittsburgh, PA 15213 USA \\ \\
Santa Fe Institute \\
1399 Hyde Park Road \\
Santa Fe, NM 87501 USA \\
  \texttt{sdedeo@andrew.cmu.edu}}

\begin{document}
\maketitle

\begin{abstract}
A wide variety of cultural practices take the form of ``tacit'' knowledge, where the rules and principles are neither obvious to an observer, nor known explicitly by the practitioners. This poses a problem for cultural evolution: if beginners cannot simply imitate experts, and experts cannot simply say or demonstrate what they are doing, how can tacit knowledge pass from generation to generation? We present a domain-general model of ``tacit teaching'', that shows how high-fidelity transmission of tacit knowledge is possible. It applies in cases where the underlying features of the practice are subject to interacting and competing constraints, as is expected both in embodied and in social practices. Our model makes predictions for key features of the teaching process. It predicts a tell-tale distribution of teaching outcomes: some students will be nearly perfect performers while others receiving the same instruction will be disastrously bad. This differs from most mainstream cultural evolution models centered on high-fidelity transmission with minimal copying errors, which lead to a much narrower distribution where students are mostly equally mediocre. The model also predicts generic features of the cultural evolution of tacit knowledge. The evolution of tacit knowledge is expected to be bursty, with long periods of stability interspersed with brief periods of dramatic change, and where tacit knowledge, once lost, becomes essentially impossible to recover.
 \vspace{0.1cm}
{\bf Significance Statement}. Many cultural practices are somewhat mysterious, even to the people that possess them. Horse riding, playing the violin, or hunting are all examples of ``tacit'' knowledge, where even experts will struggle to articulate the majority of what goes into successful performance. These are in contrast to more formal systems, such as those that can be implemented by a procedural programming language. Tacit practices are equally opaque to newcomers, and no amount of watching from the sidelines will enable them to perform like an expert. Despite these obstacles, tacit practices are found across the cultural and historical record, and are reliably taught and passed down from generation to generation. We present a model that shows how this can happen. Under natural assumptions about how different features of the practice fit together, it becomes possible for a teacher, with precise but minimal intervention, to guide a student in such a way that the full practice ``locks in'' with neither teacher nor student being aware of details. Among other things, our model has implications for how traditional practices can sustain themselves, evolve, and be lost over long timescales.

\end{abstract}


\maketitle

\clearpage

Tacit knowledge is ``what we know but cannot say''~\cite{tacit_p}: the vast array of complex cultural practices whose principles cannot be verbalized. The ``tacit dimension'' goes by many names, including ``working knowledge''~\cite{harper1987working}, ``practical'' knowledge~\cite{archer2000being}, ``know-how''~\cite{ryle2009concept}, and ``knowing-how''~\cite{harris2007ways,sep-knowledge-how}. Tacit knowledge is found in everything from sports~\cite{jakubowska2017skill, nyberg2014exploring} and artistic performance~\cite{kaastra2016tacit} to architecture~\cite{alexander1977pattern}, medicine~\cite{patel1999expertise}, and science itself~\cite{Brock2017}, and it is seen in contexts ranging from traditional crafts~\cite{marchand2008muscles} to the professions~\cite{sternberg1999tacit} and organizations \cite{baumard1999tacit} of the modern world.


Just as with any other form of culture, tacit knowledge must be transmitted from one generation to the next. In cultural evolution, standard transmission mechanisms include teaching (where a teacher communicates their understanding to a learner), emulation (copying an end product), and imitation (copying the actions that produce the product); see~\citeA{tomasello1993cultural, caldwell2009social, hoppitt2013social, morgan2015experimental}. While these three mechanisms can account for part of how culture is transmitted, they struggle to explain the case of tacit knowledge. Three aspects, in particular, make the task challenging.

First, tacit knowledge is a mental representation. To be transmitted, that representation must be in some way made public~\cite{sperber2007culture}. One main way to do so is verbal instruction, and a great deal of culture is passed down by speech alone~\cite{morgan2015experimental, bietti2019cultural}. However, tacit knowledge cannot be transmitted in this fashion~\cite{neuweg2004tacit, collins2010tacit}, because, by definition, even those who have the knowledge would not know what to say. 

Second, tacit knowledge is combinatorially complex. It provides those who possess it with a set of contingently deployed, interconnected skills ~\cite{stout2002skill,seifert2013key}. Constitutive aspects of an expert's tacit knowledge may become relevant so rarely---say, ``under pressure'', or in an exceptional context---that even the most diligent student may never encounter them through observation alone. This makes it difficult for a standard alternative to explicit instruction: the target goals, and their contingencies, are too various and mutable for straightforward imitation or emulation to work. 

Third, tacit knowledge includes knowing which aspects of behavior constitute the practice, and which are incidental. This makes imitation difficult: if a learner is to acquire skills through imitation, she needs to know, or be able to infer, what is relevant to imitate, including whether an action is understood as instrumental or not~\cite{gergely2002rational}. This knowledge, however, is itself tacit. For instance, I may be able to improve my technique by watching a skilled performer, but only after I have enough tacit knowledge to know the relevant from the incidental. A novice at the violin cannot learn by watching an orchestra perform, and imitation alone cannot sustain the cultural transmission of tacit knowledge. Similar challenges occur for emulation: when knowledge is tacit, a learner cannot determine which features of the end product matter.

This paper presents a domain-general model that shows how, despite these challenges, tacit knowledge may be faithfully transmitted. The solution we propose sees tacit knowledge as the emergent product of a network of interacting constraints, and transmission as a process of guiding a learner to a solution by the simultaneous, and mutually interfering, demands of both a teacher and the environment. The knowledge is tacit even in transmission because only an enigmatic fragment is ever present in the mind of either teacher or learner. The structure necessary to reconstruct the practice emerges from the interaction between the practitioner and the environment, and the teacher's task is to guide a learner towards the correct use of that structure. In particular, by careful intervention on a small fraction of the features, a teacher can guide the learner to discover the full structure of the culturally-specific solution.

Our model shows how only around 10\% of the task need be conveyed by a teacher's intervention. This helps make sense of a key feature of teaching seen across the anthropological record, where the most common forms of teaching in the cultural record are low-cost and involve significant underspecification; see, e.g., \citeA{kline2013teaching}. This is, of course, in contrast to the ``Western'', or WEIRD~\cite{henrich2010weirdest}, image of teaching as rationalized, explicit, and high-cost.

The results can also, as we show, explain a puzzling feature of cultural evolution: the fact that culture appears to proceed in a bursty fashion, with long periods of stasis interspersed with short bursts of chaotic innovation leading to rapid and dramatic changes. Bursty evolution is common in cultural evolution (see, \emph{e.g.}, \cite{PhysRevX}). It is also a defining feature of prehistory: bursty changes in material culture, for example, provide the basis for how we divide prehistoric cultures into distinct periods. 

We present this solution in three parts. We first present the model, showing how the mental representation of the practice is embedded in a network of embodied constraints, and how a teacher intervenes to help construct the representation for a learner. We then show how the fragmentary nature of these interventions combines with the environment to allow for the faithful transmission of knowledge from generation to generation. Finally, we present our results on the longer cultural dynamics which arises spontaneously from the model, namely bursty evolution.


\section{Model}


\begin{figure}
    \centering
    \includegraphics[width=\linewidth]{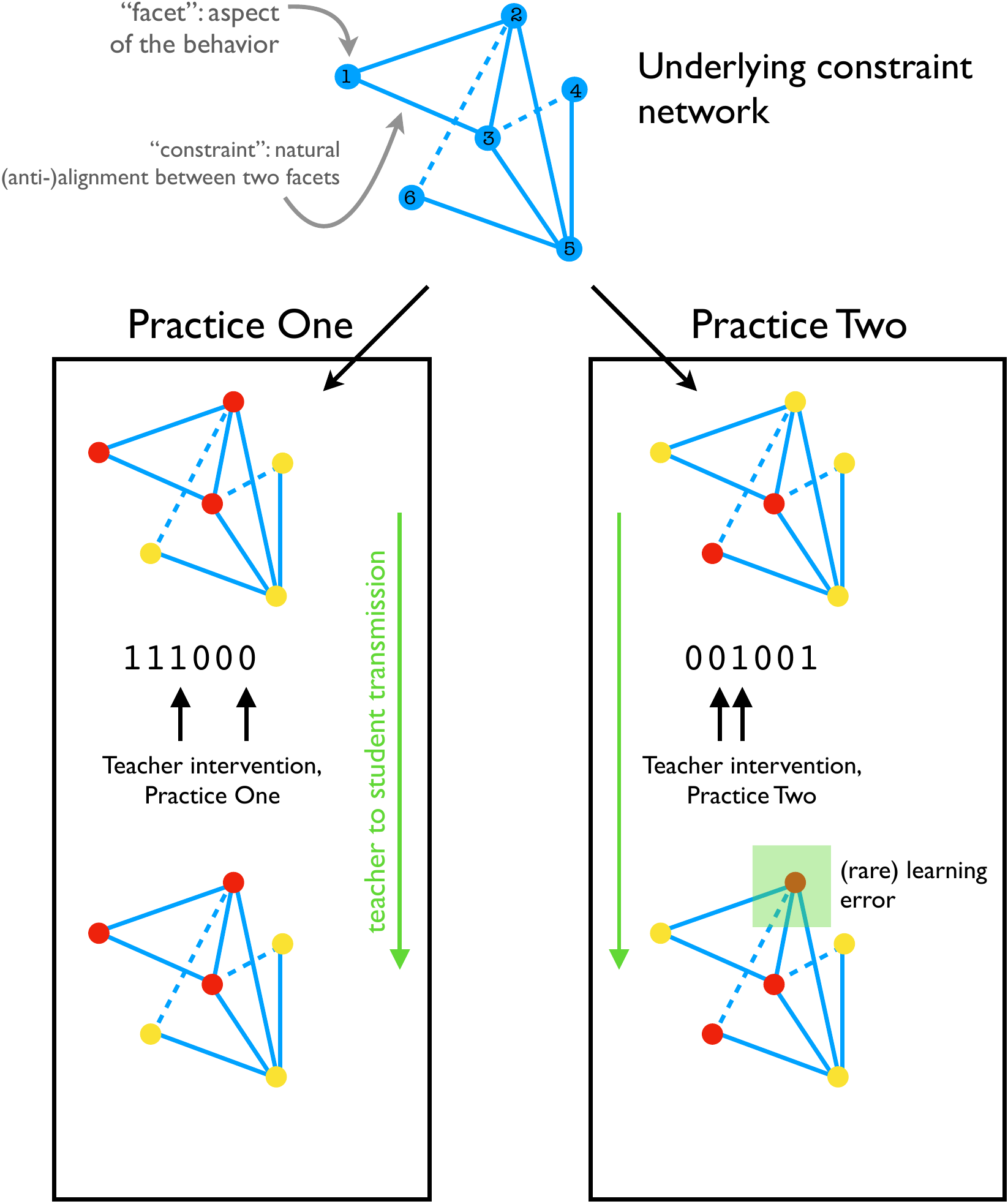} 
    \caption{A constraint model of tacit knowledge transmission. A tacit cultural practice is a (usually partial) solution to a complex network of interacting constraints between aspects (``facets'') of the problem; we show a simple toy example here with only six facets, each of which takes on a binary value, \emph{i.e.}, an agent performs any particular aspect of the task in one of two mutually exclusive ways (red or yellow; written as one or zero, respectively). Constraints are pairwise, preferring either alignment (solid lines), or anti-alignment (dashed lines), of the facets.} 
    \label{tacit_picture}
\end{figure}
Tacit knowledge can appear in a wide variety of domains. Our model attempts to capture the generalizable features of tacit knowledge by working at an abstract level. Our model shares a number of characteristics with that of~\citeA{alexander1964notes}; in particular, both describe different forms of tacit cultural knowledge as systems of constrained and interacting choices. The particular mathematical instantiation of our model is closely related to the Boltzmann machine paradigm common in machine learning~\cite{ackley1985learning}, Hopfield networks used in neuroscience~\cite{hopfield1982neural}, and the ``spin glass''  systems studied in physics~\cite{sherrington1975solvable}.

A particular case of tacit knowledge, here, is defined as a list of conditional behaviors. We refer to these as ``facets''.  As an example, consider horse-riding. A particular style of riding corresponds to a tacit knowledge practice, and each style will involve a complex relationship between how, for example, the rider places their limbs in response to the movements of their mount. 

In principle, each of these conditional behaviors can be specified by one of a set of symbols. If the body is in position so-and-so, and the horse does so-and-so, should the rider respond by lowering their hands (``option $A$ for facet one'', or $A_1$ for short), or, alternatively, by raising them (option $B_1$)? Should they relax their back (option $A_2$) or straighten it (option $B_2$)? Part of the facet specification for one style of riding might be $A_1A_2$, while another style might be $B_1B_2$, and so forth. The number of facets is potentially very high.

The second step of our model considers the interacting constraints between these different facets. What a rider does with one part of her body in a particular context will, because of the nature of the human or equine body, or because of the particular artefacts used for riding (the saddle, tack, and so forth), be more or less consonant with what she does with another part of her body. For example, the combination of lowering one's hands and relaxing one's back may be a particularly consonant combination, while lowering one's hands and straightening one's back may not---\emph{i.e.}, an ``incorrect'', or inexpert, response might be $A_1B_2$. A good combination will be something that, all other things being equal, the person can receive some sort of feedback on from the environment. For example, a consonant pairing may take less effort, or provide some other noticeable benefit such as fluency.

These consonance relationships, taken together, are called the constraint network. In our model, each facet in the network bears some relationship to the others. This can be a direct link, as in the example above, or it can be an indirect link, mediated by intermediate facets. For example, we might imagine a third conditional behavior with two possibilities, $A_3$ and $B_3$, and that $A_3$ is more consonant with $A_2$, and $B_3$ is more consonant with $B_2$. The choice for the third conditional behavior, in other words, is influenced by the choice for the second conditional behavior. Because, however, the choice of $A_1$ vs $B_1$ is in turn influenced by the choice of $A_2$ vs $B_2$, the choice of $A_3$ vs $B_3$ has an impact as well. Such a network of interactions operationalizes intuitions of what makes a practice coherent. 

The simplest version of such a model takes each facet to be a choice between one of two options ($A$ or $B$; or more simply $0$ or $1$), and for the interactions between facets to be pairwise only. An example of such a network is shown in blue at the top of Fig.~\ref{tacit_picture}, with the very simple case of six facets. Each node corresponds to a facet, and lines between nodes reflect the two different types of consonance relationship. A solid line (\emph{e.g.}, the one connecting facets 1 and 2) says that the two facets in the ``same'' state are preferred, while a dashed line (\emph{e.g.}, the one connecting facets 2 and 6) says that the two facets prefer to be in the opposite state. Thus, for example, the setting $A_1A_2$ is preferred to $A_1B_2$ (1 and 2 in the same state), all other things being equal, while $A_2B_6$ is preferred to $A_2A_6$. For simplicity we can write out the full specification of the system as a binary string. One example of a string that satisfies many, though not all, of the constraints, is $111000$; in this case, among other things, it satisfies the constraint that aligns facets 1 and 2, and that anti-aligns the facets 2 and 6.

Networks of interacting constraints like these, that include both preferences for alignment and anti-alignment, are often difficult, if impossible, to satisfy. In our simple example, facets 2, 5 and 6 cannot be set in a way that satisfies all three constraints simultaneously---as can be verified by trying the different combinations. Any particular specification for the facets, in other words, leads to difficulties. 

Generically, there are different ways to satisfy these competing demands. Some specifications are better than others, and some are worse, but in general any particular tacit knowledge practice is a matter of how these difficulties are navigated. ``Practice One'' in our figure, for example, violates two constraints, while ``Practice Two'' violates four. Others, not shown, are much worse; for example, the practice $101100$ violates six constraints.

When a practice is a reasonably good solution to the constraint network, a practitioner who has learned the practice finds it easy to maintain. Deviations from the standard in many facets can be sensed and corrected. Consider, for example, someone implementing Practice One. If she deviates by switching from the ``0'' state to the ``1'' state in facet four, she will experience an increased level of negative feedback from the environment, since she is now aligning with facet three (when it is more consonant to anti-align), and anti-aligning with facet five (when it is more consonant to align). This provides her with a signal that can be used to return to the standard. Even if she is unaware of which facet deviated, she can make little (\emph{i.e.}, roughly single-facet) adjustments in her behavior until consonance returns. When the practice is a reasonably good solution, in other words, the practitioner only needs to implement the solution. She does not need to understand it. Stable solutions like these are candidates for culturally transmitted tacit knowledge practices.

Transmission of the practice is now a matter of guidance. If the learner can be guided by a teacher close enough to the standard practice, the feedback from constraints will be sufficient to maintain her there. A very simple model of guidance is the intervention of the teacher to fix some of the facets into the culture's pattern. These may include physical interventions (to teach fly fishing, for example, a novice may be guided in proper form by literally tying his wrist to the rod), scaffolding (use of the barre in ballet), mnemonics (``eye on the ball'', which maintains proper stance in tennis), or simple verbal guidance from the teacher (``back straight!'').

Careful interventions can do a great deal. In our toy example, practice one can be efficiently transmitted to the next generation by fixing only two critical nodes (nodes three and six). A learner who obeys her teacher's guidance in these two facets can learn the full pattern simply by remaining attentive to environmental feedback. She need only minimize the number of violated constraints, subject to the two instructional demands.

That effective subset of interventions (a kernel), when placed in an embodied context, reliably activates the characteristic and flexible behaviors of an expert.  The very nature of tacit knowledge means that the teacher is unaware of the exact nature of practice she exemplifies. However, the structure of the problem also can enable ``tacit teaching'', where the teacher intervenes in a fraction of the facets but nonetheless passes on the practice to some of the learners with near-perfect fidelity. 

\section{Results}

In order to study this model quantitatively, we need to specify how a learner responds to the constraint network. We choose a general specification, known as the maximum entropy model, which fixes only the average correlation between nodes with direct constraints. Once the constraint network is specified, this model has a single free parameter, $\beta$, which governs the learner's sensitivity to constraints. When $\beta$ is very low, the learner pays little attention to the constraints of her environment; when $\beta$ is very high, she is exceptionally rigorous. Assuming that the teacher is obeyed rigorously enough, our results are not particularly sensitive to the value of $\beta$, as long as it is past a critical point---essentially, the learner needs to be reasonably attentive to her environment, for some notion of reasonable. Mathematical details are available in the Materials and Methods. 

\subsection{Tacit Teaching}

We first consider tacit teaching itself. Given a particular constraint network and cultural practice, we consider the effect of different kernel sizes (see Materials and Methods) on the fidelity of transmission. Fidelity is measured by Hamming distance, which counts the number of facets in which the student differs from the teacher. A Hamming distance of zero indicates perfect transmission.

The results of our simulations suggest that under a variety of conditions, perfect transmission can be reliably obtained even when the number of interventions is significantly smaller than the number of facets. The kernel needs only be a tiny fraction of the whole, and a skilled teacher in possession of that kernel would still be able to transmit the whole practice to the learner, even if only a small amount of information is conveyed between them.


\begin{figure}
    \centering
    \includegraphics[width=\linewidth]{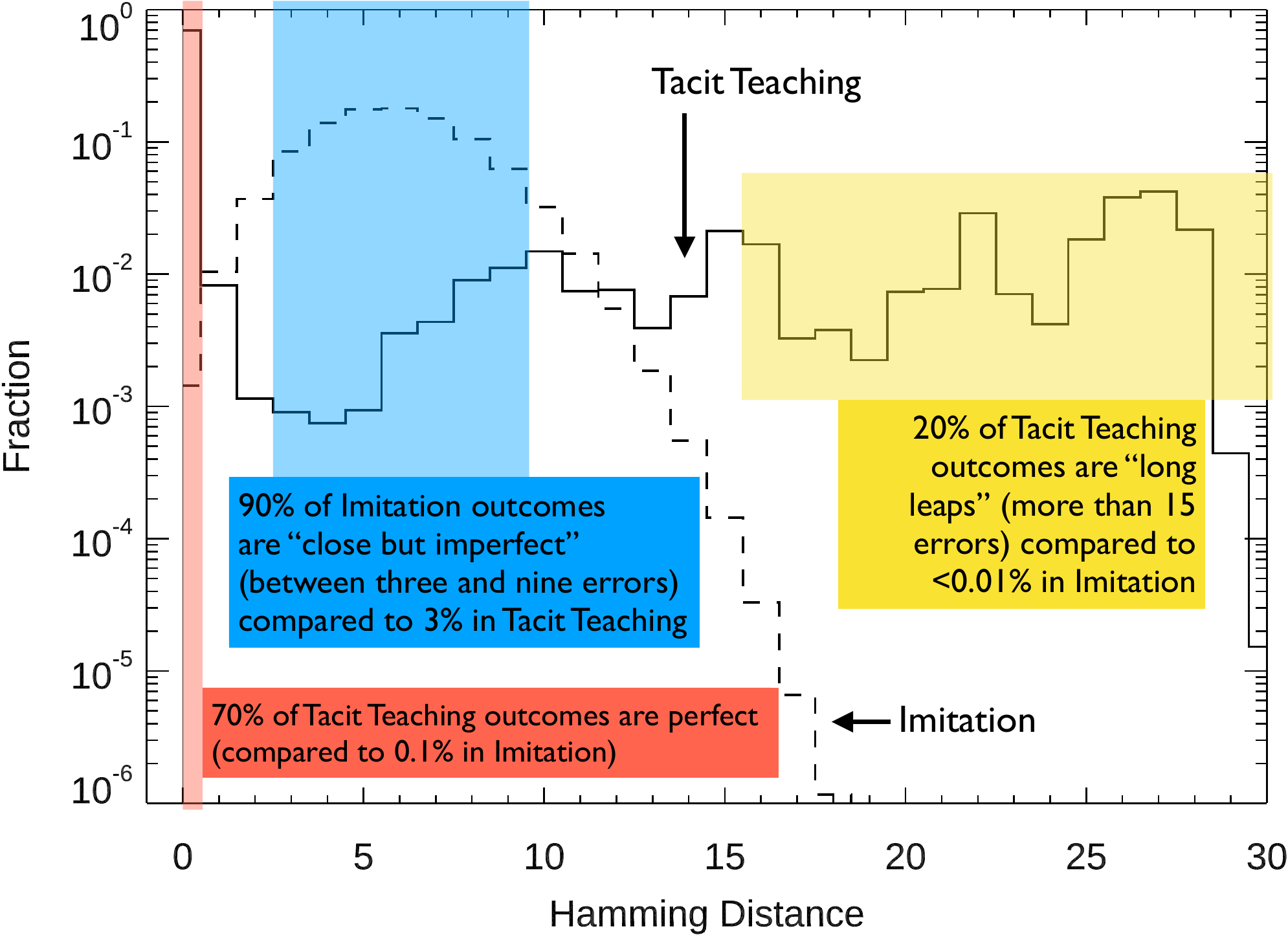}
    \caption{Tacit teaching leads to high-fidelity transmission from teacher to student. Shown here is the distribution of learning outcomes for the tacit teaching case (solid line) for the test case of 30 facets and (only) four teacher interventions. Even though the teacher intervenes in less than 15 percent of the facets, perfect reproduction occurs approximately 70\% of the time (see the extreme left side of the distribution). The students who fail to learn properly, by contrast, often end up with very different solutions---the distribution of errors is non-normal (non-binomial) and long leaps are just as, if not more, common as small deviations. A null ``Imitation'' (or ``copy error'') model (dashed line) has very different properties. If it is to achieve the same average accuracy as the tacit model, it must sacrifice any hope of faithful transmission.} 
    \label{tacit_teaching}
\end{figure}
An example of this phenomenon is shown in Fig.~\ref{tacit_teaching}. In this case, a thirty-facet practice can be transmitted with very high fidelity by intervening in only \emph{four} facets. Despite the fact that, on the surface, only a tiny fraction of the total information is conveyed between teacher and student, perfect fidelity can be achieved nearly 70\% of the time.

Neither teacher nor learner need know, in any conscious fashion, the correct pattern in all thirty facets---indeed, they need not even know how many facets there are. All that is needed for effective transmission is (1) that the teacher keep in mind four key features of the learner's behavior, and (2) that the learner attend to the teacher's guidance while remaining attentive to the consonance demands of her environment. This is not only sufficient to guide the learner to the full thirty-facet practice, but also to avoid other, potentially tempting---\emph{i.e.}, stable and similarly optimal---solutions that can be thought of as alternative cultural practices.

Two things are evident from Figure 2. First, as noted, a majority of students learn the practice exactly. If teachers for the next generation are drawn from this sub-population, the practice can persist with high levels of fidelity for multiple generations. Second, the distribution of errors is highly non-normal; of those who fail to learn, there are just as many who learn a practice (say) three Hamming units away as twenty. Poor transmission is therefore expected to be far more noticeable.

This distribution arises because the underlying constraint network serves to correlate the errors made in learning: informally, a failed student learns ``bad habits'' that connect together and re-enforce each other, driving the learner into a totally different part of solution space. This space is usually less optimal than the correct answer, but may have at least a modicum of stability. A simple example is in the teaching of juggling. A minority of learners find a satisfying, but in the end suboptimal, solution to the problem of juggling two balls that involves passing, rather than tossing, one of the balls from one hand to the other.

Once a student has learned enough of these bad habits, further teaching may be in vain. Matching the teacher's practice would now require shifting a large number of facets simultaneously. The only other solution to this problem is if the student can start again---in our simple model, the necessary ``beginner's mind'' is a random choice for each facet---and pay greater attention.

A number of consequences flow from this distribution of errors. First, it is easy to spot the majority of students who fail to imitate the practice: their overall behavioral pattern is generally very different from the cultural norm and (furthermore) the practice they do adopt is expected to be generally less effective in (for example) competition with learners who have correctly grasped the norm. 

Second, however, not all errors are a combination of bad habits. It is entirely possible that a small fraction of students who fail to learn achieve, instead, "true" alternative practices, meaning solutions to the constraint network that are, if not equally good as the standard practice of their culture, would be at least similarly stable. 

This leads to an interesting paradox. On the one hand, tacit teaching is, despite the fragmentary nature of the teacher's interventions, extraordinarily reliable. A majority of students learn the practice faithfully. On the other, however, tacit teaching is also highly evolvable. The deviations that do occur are often significantly different from the standard practice.

One way to understand this result is to compare it to a null model, an imitation or ``copy error'' model. This model assumes that all of the facets are observed by the student and copied independently with some level of error. To compare the copy error model to tacit teaching, we tune the error rate of copying so that the average Hamming distance matches that of the tacit model. The imitation/copy error model is shown in Fig.~\ref{tacit_teaching} by a dashed line. 

When comparing the two error distributions, two things stand out. On the one hand, the imitation/copy error model achieves basically zero fidelity: it is essentially impossible for a learner to match the teacher's practice, despite the assumption that he is aware of and can attend to all of the facets in turn. Second, despite this high error rate, it is also very hard for the copy error model to make long leaps and discover viable alternative practices. The vast majority of outcomes for imitation lead to ``close but imperfect'' outcomes, with an error rate of 1/6th; only around 0.1\% of learners reproduce the practice perfectly, and less than 0.01\% produce long leaps that modify more than half of the practice. 

On the other hand, the tacit model produces a spectrum, with a large number of perfectly accurate students, and a small number of outlier eccentrics. Most of the outliers, of course, fail to create a new practice, but a small number may find novel, but stable and teachable, solutions. This has suggestive consequences for cultural evolution dynamics, especially with regards to diversity and evolvability. We examine them in the next section. 

\subsection{Population Level Dynamics}

No teaching method is perfect, and every culture needs to deal with the fact that some fraction of the students will fail to learn. While the tacit teaching model can achieve high fidelity, not everyone is successful. If transmission is always a matter of independent learners who each become teachers to a new group of their own in turn, the practice will soon decay. 

One institutional solution to this problem is for the learners in each generation to agree on a consensus practice that is taught to the next. If there are ten students, for example, in our thirty-facet model above, roughly seven of them will learn the same practice. If consensus is simply a matter of voting on which pattern (or, rather, kernel) will be taught to the next generation, then error-free transmission can be sustained over many generations. This is robust for two reasons: because, on average, we expect the standard practice to dominate, but also because the deviations are often idiosyncratic. Even if the standard practice does not obtain a majority, it will usually retain a plurality.

Not always, however. This is in part because idiosyncratic fluctuations are not random: ``bad habits'' tend to drive students to the same, suboptimal parts of the solution space. This means that it is not that difficult or rare for a non-standard practice to obtain a plurality. When this does happen, two things follow. First, the initial, standard, culture's practice is lost. Second, it is replaced by something that is usually suboptimal compared to the original.

Suboptimal solutions, in turn, are more difficult to learn because there are more nearby solutions that are equally good. A learner who deviates in one or two facets may find that, rather than upset a fine balance, she has satisfied just as many constraints as she did before. Now there is no good signal to lead her back to the original pattern, and, unless the teacher makes more interventions, transmission will be unsuccessful.

Taken together, these effects predict that the cultural evolution of tacit knowledge is \emph{bursty}. Long periods of stability, in which cultural practices change very little, are interspersed with chaotic periods. These chaotic periods begin with a long leap in the solution space, and the original tradition is completely lost. Communities of practice in these chaotic periods are then much worse at preserving their (new) traditions, and make long leaps in turn. This continues until a new, sufficiently stable, practice is discovered. A longer period of high-fidelity transmission commences, and the cycle repeats.

\begin{figure}
    \includegraphics[width=0.85\linewidth]{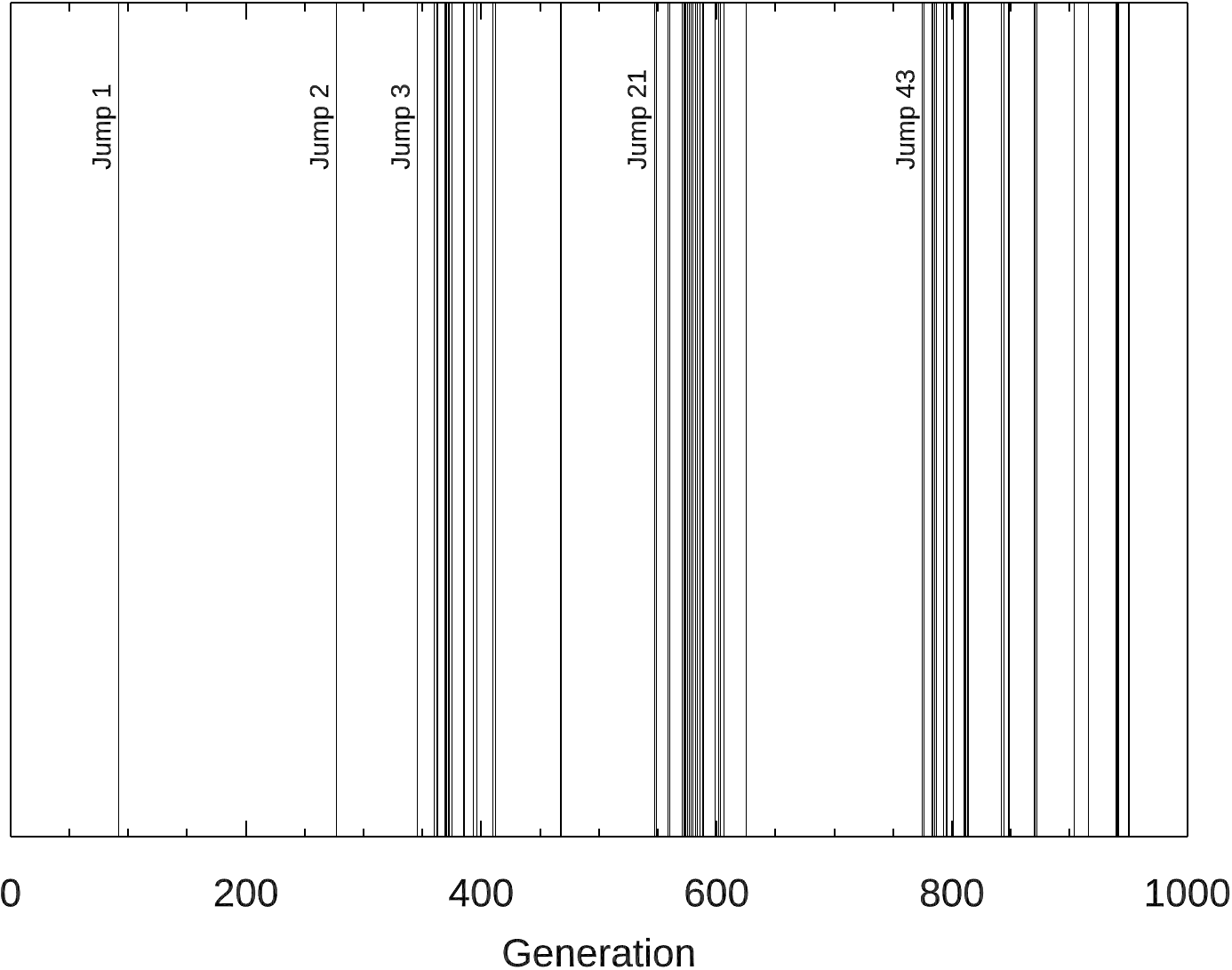} 
\caption{A sample inheritance sequence of 1000 generations, showing how jumps from one practice to another tend to be concentrated in time. The $x$ axis labels the generation number; a vertical line indicates a jump from one practice to another. Lines are clustered in groups indicate chaotic periods. Results of a simulation with thirty facets and four teaching nodes. \label{lines}}
\end{figure}

\begin{figure}
    \includegraphics[width=0.95\linewidth]{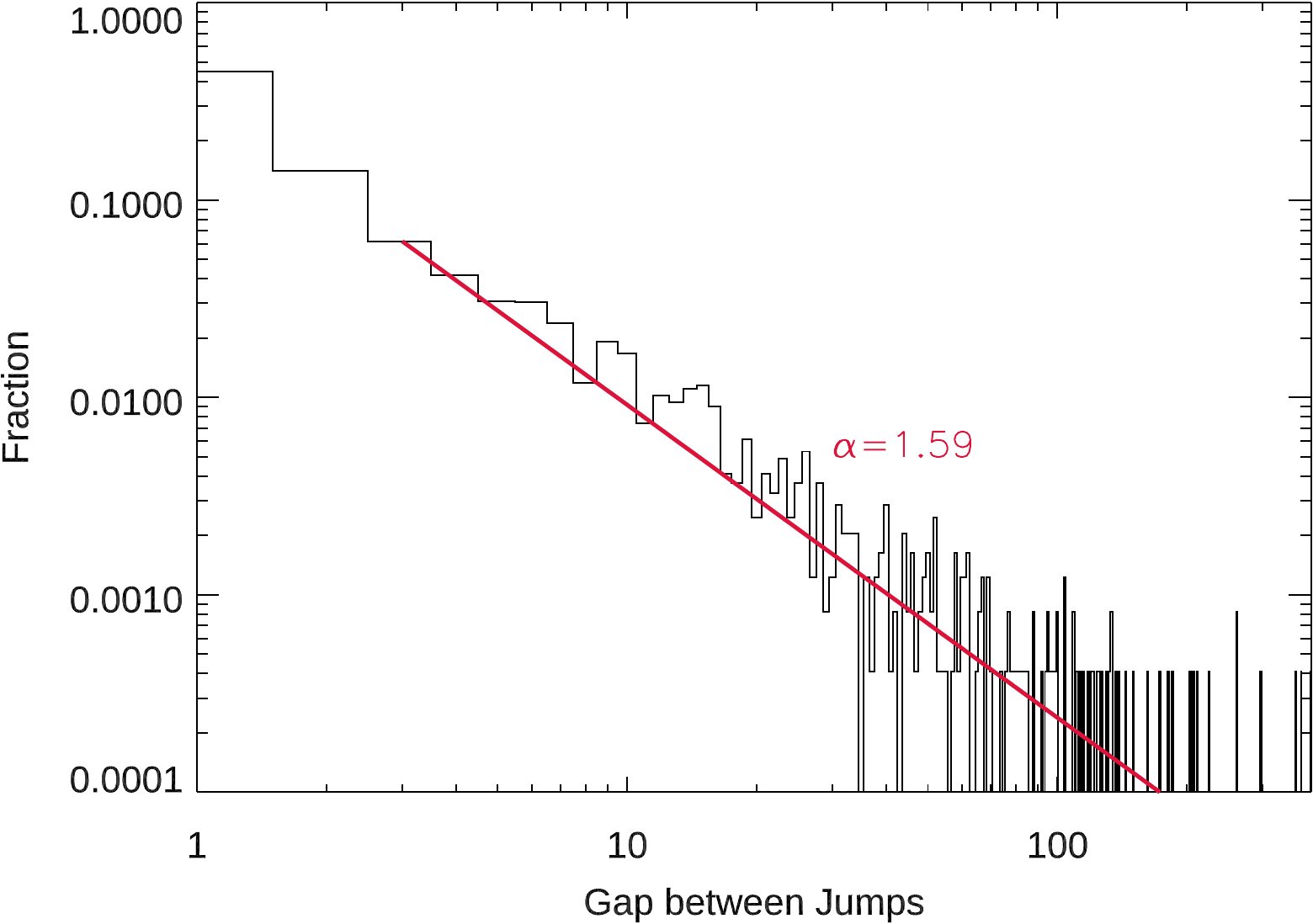} 
\caption{The distribution of gaps between jumps, in sample simulations with the same parameters as Fig.~\ref{lines}. The majority of gaps are very short, corresponding to the highly turbulent periods visible in the simulation of the Fig.~\ref{lines}. Standard analysis methods~\protect\cite{clauset2009power}, applied to the tail of the data, show that it follows a scale-free power law with index approximately $1.6$. \label{dist}}
\end{figure}

This is shown, first, in Fig.~\ref{lines}, with a sample simulation of seven learners conducting a majority vote. A vertical line indicates a generation where the practice has switched. While the first two jumps are isolated events, the third jump leads to a chaotic cascade of jumps in the next fifty generations; similar turbulent periods appear every few hundred generations. Fig.~\ref{dist} shows the distributions of gaps between jumps. This has a roughly power-law, or scale-free, distribution characteristic of turbulence. The majority of jumps are followed, one or two steps later, by another jump; once in a long while, however, these rapid jumps are interrupted by many hundreds of generations of stability. Because this distribution is a power law, there is no characteristic limit for how long this stability can last.

A final way to visualize this bursty behavior is to track the evolution of the practice itself. Practices, as we have conceived of them, are high-dimensional objects; a thirty-facet practice lies on one of the vertices of a thirty-dimensional hypercube. This is, of course, impossible to visualize. However, we can use the fact that stable practices are sparsely distributed to our advantage. Since only a small fraction of the solution space corresponds to stable practices, we can use a dimensionality-reduction algorithm to map the shifts from generation to generation onto the two-dimensional page.

\begin{figure}
    \centering
    \includegraphics[width=\linewidth]{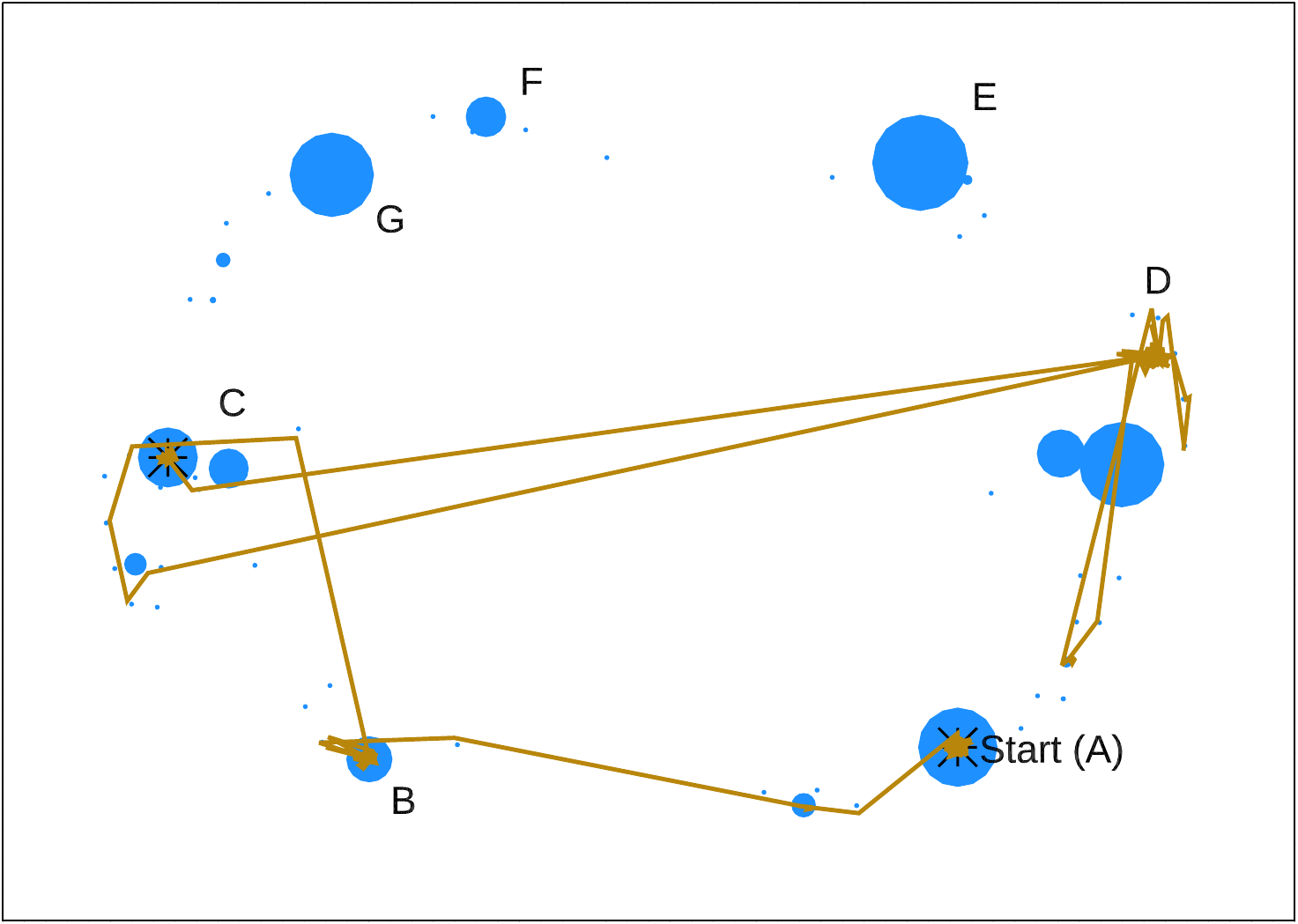}
    \caption{Exploration in cultural space, visualized in a sample simulation with the same parameters as Fig.~\ref{lines}. While any particular practice is a thirty-dimensional binary vector, an approximate visualization is possible with the MDS algorithm. Circles correspond to different practices, with size proportional to stability. The yellow line follows the trajectory of a simulated culture, which begins in the practice labelled A, and wanders, in a characteristically bursty fashion, to land, finally, in practice C. \label{awesome}}
\end{figure}

This is shown in Fig.~\ref{awesome}. Each blue circle represents a point on the hypercube of tacit practices. The two-dimensional layout, provided by the MDS visualization algorithm, approximates Hamming distance: circles that are nearby each other on this plot have more facets set to the same value. Circle size is proportional to stability; larger circles indicate solutions that both satisfy more of the underlying constraints and are stable under perturbation. In this model, with thirty facets, roughly a dozen distinct (significantly) stable practices can be found.

The yellow line shows a sample evolutionary trajectory through this space. The simulated population begins in the relatively stable and teachable practice A, which it maintains for a long time. After a hundred generations or so, it makes a long jump to practice B. Practice B is less teachable (it is rarely transmitted faithfully from teachers to learners), and the culture enters a period of instability, making additional long jumps to practices near practice C, and spending tens of generations in the a hard-to-maintain cluster of practices near practice D. Eventually the system returns to, and settles down in, the highly stable practice C. Other practices (E, F, G, etc) remain undiscovered by this culture even after many thousands of generations.

For simplicity of presentation, our discussion has focused on networks with thirty facets and with couplings between any two facets drawn from a uniform distribution between one and negative one (a ``random network'' model; see Materials and Methods). Simulations of both larger and smaller networks, and networks with different topologies, produce essentially identical qualitative features, at both the population and the individual level.

We find that larger networks can support more cultural practices, and tacit teaching requires more interventions as the complexity of the practice grows; for networks between ten and one hundred nodes, we find that tacit teaching with majority accuracy (\emph{i.e.}, at least half the time, a randomly chosen student matches the practice exactly) requires interventions on around 10\% to 15\% of the facets. This linear scaling is preserved for a variety of different distributions of edge weights.

Modelling the system of facet constraints as a random network has limitations. In many situations, we expect facets to organize themselves into roughly distinct ``modules'' with tight interconnections within each module, and fewer, more disorganized connections, between modules. 

These modular organizations are expected under a range of circumstances. For example, a modular organization is expected when the facets concern material properties of the task, where spatial and temporal separations can generate nested topologies. For example, when behavioral facets include the relative positions of different parts of the body, we expect there to be tighter constraints between groups of muscles that connect to the same joints. We also expect the emergence of modularity under generic tinkering and bricolage processes, as originally described by \cite{alexander1964notes}; more recent work suggests that, if the underlying constraints are built up by combining and repurposing earlier practices, the resulting network will have high levels of modularity~\cite{sole2020evolving}

Under the assumption that each module has only two configurations, our results now apply at the module level. To teach a practice with thirty modules, for example, we expect tacit teaching to require interventions in (roughly) four of them. Modules may be more complex, meaning that they may be able to support more than two internal configurations. A simple information-theoretic argument suggests that this will scale logarithmically in the number of additional interventions. If each module has $N$ solutions, for example, then the demands on tacit teaching increase, albeit slowly, by a factor of $\log_2{N}$.


\section{Discussion} 




Cultural evolution, as a field, has often opted to ``black box'' how information is socially transmitted and learned~\cite{heyes2016blackboxing}. Models such as~\citeA{henrich2004demography, boyd1988culture, mesoudi2011variable} give us great insights into cultural dynamics at the population level. However, they have done so in part at the cost of ignoring the complexity of the interactions between cognitive agents through which culture is acquired and transmitted. 


Our model provides an explicit and quantitative account of the relationship between teacher and student in the commonly-encountered case of tacit knowledge. It shows how high-fidelity ``tacit teaching'' is possible, even in the case where both teacher and student lack conscious knowledge of up to 90\% of the components of the practices. A small amount of guidance, well-presented, allows the majority of students to ``lock in'' an efficient, culturally-widespread practice. This is possible only when the features of underlying practice are subject to specific constraints and echoes observation of skill acquisition dynamics in ecological contexts ~\cite{button2020dynamics, hristovski2006boxers}.

Our results make empirical predictions for cognition at the individual level. One key feature of tacit teaching is the presence of an unusual and non-exponential distribution of learning errors: when tacit teaching is in place, we expect even diligent learners to, occasionally, learn something that diverges significantly from the correct performance. Conversely, we expect to find a deficit of near misses: students who are close to getting it, but miss only in one or two aspects. More generally, as seen in Fig.~\ref{tacit_teaching}, we expect a characteristic pattern of error-making that looks very different from a model where the teacher teaches everything, and the student learns each piece independently. When most students do extremely well, but a small fraction, with otherwise equivalent abilities, do extremely poorly, it may be a sign that tacit knowledge is at play.

These results have, in turn, implications for cultural evolution. They predict bursty, and sometimes very long-leap, innovations, with a heavy-tailed power law distribution that makes it possible for a practice to change without going through a series of gradual mutations. These long leaps can enable, potentially, rapid adaptation to new condition (\emph{e.g.}, changes in the underlying constraint network). They come at a cost, however: once a leap has been made, it is very difficult to recover the prior practice, except by accident.

\section{Conclusions}

A cultural tradition is more than just a list of behavioral features. It is enabled by how those features fit together into a larger logic dictated by mental, material, and environmental constraints. We have presented a minimal model that allows us to capture this higher-order logic, and to thereby go beyond accounts of cultural evolution that focus on the acquisition of particular traits.

Attending to the cognitive aspects of transmission has great benefits. It reveals how these interactions cannot only channel culture practices, but make it possible to faithfully transmit them, from generation to generation across long periods of time, with a teacher's intervention serving as a seed for what will eventually be the learner's full practice. It also shows how this kind of high-fidelity transmission can coexist with the dynamics and long-leap changes that characterise the macroevolution of culture.


\clearpage

\section{Materials and Methods}

Our goal is to model how the interaction between facets leads to distinct tacit knowledge practices, and the ways in which (highly-partial) teaching can allow a practice to be passed to a new student. In order to do this, we adapt an approach used in a variety of cognitive models known as the maximum-entropy principle~\cite{schneidman2006weak,lezon2006using,seno2008maximum,bialek2012statistical,lee2015statistical}. Our model makes minimal assumptions about how facets interact. In particular, in the absence of teaching, the model fixes only the pairwise correlations between each pair of facets. Meanwhile, effect of teaching is to fix the average value of the particular facet being taught. 

All the relevant properties of the teaching process can be captured once we can compute the probability distribution of the learner over the different facet patterns. Following the discussion above, we assume that each facet for the learner, $\sigma_i$, can take on only one of two values; for simplicity, the two choices can be represented as $+1$ and $-1$. Then the probability distribution under the minimal model can be written~\cite{schneidman2006weak} as
\begin{equation}
P(\{\sigma_i\}) = \frac{\exp{\left( \beta\sum_{i,j} r_{ij} \sigma_i\sigma_j + \tau \sum_{i\in T} t_i \sigma_i\right)}}{Z},
\label{awesome_eq}
\end{equation}
where $r_{ij}$ is a matrix that describes the coupling between facets $i$ and $j$; a positive value of $r_{ij}$ indicates a preference for the two facets to be in the same state, and a negative value for them to be in opposite states (corresponding to the solid and dotted lines in Fig.~\ref{tacit_picture}, respectively). $T$ is the set of nodes that are taught by intervention (the nodes marked with arrows in Fig.~\ref{tacit_picture}), $t_i$ is the teacher's intervention (either $+1$, indicating a preference for the ``positive'' practice, or $-1$, indicating a preference for the negative practice). The two constants $\beta$ and $\tau$ govern the strength of the interaction between facets, and the influence of the teacher, respectively. 

Finally, $Z$ represents the normalization constant for the distribution, however we do not need to calculate this explicitly, because we can estimate $P(\sigma_i)$ by starting the system in a random configuration and simulating it forward dynamically. This can be done using well-known Monte Carlo techniques (we use Glauber dynamics~\cite{glauber1963time}); intuitively, this method determines the most likely configuration for the system after the student has interacted with the teacher for sufficient time. Formally, and within the context of this model, ``sufficient'' corresponds to what is know as ``burn-in time''---the number of iterations of the interaction necessary for the learner's pattern to decorrelate from the (random) initial conditions they begin in. 

Eq.~\ref{awesome_eq} appears in many models in machine learning (Boltzmann machines~\cite{ackley1985learning}), neuroscience (Hopfield networks~\cite{hopfield1982neural}), and physics (spin glass models~\cite{sherrington1975solvable}). The salient feature of all of these systems is the existence of multiple, distinct, ``metastable'' (\emph{i.e.}, long-lived) patterns of activation. In the Hopfield case, these correspond to different ``memories''; for us, they correspond to different practices. The goal of the teacher is to guide the learner to her same solution.

(A terminology note: in the physics-style framing where these models first appeared, $\beta$ and $\tau$ are sometimes known as ``inverse temperatures''. When $\beta$ is small, for example, this corresponds to a ``high temperature'' system where facets fluctuate largely independently of each other and the value of any particular facet is largely uncorrelated even with those it is supposedly constrained by. Lowering $\beta$ corresponds to increasing the influence that different facets have on each other. Similarly for $\tau$, which quantifies the strength of interaction between the teacher's application of a facet choice, and the resulting student performance.)

We simulated different interaction configurations, where $r_{ij}$ in any particular simulation is drawn from a random distribution, uniform between $-1$ and $+1$ (our results are insensitive to the precise nature of this distribution). Once the interactions are fixed, the key parameters are $\beta$ and $\tau$. We set $\tau$ to be much larger than unity (in practice, ten), and $t_i$ to be zero for the facets $i$ that the teacher does not intervene on, indicating that the teacher can make a strict intervention; \emph{i.e.}, that she can fix that small number of the student's facets with near perfection. Meanwhile, $\beta$ indicates the extent to which the student is sensitive to the interacting constraints of the facets themselves. As discussed above, when $\beta$ is zero, for example, the facets that are not being taught are completely free and uncorrelated. When $\beta$ is very high, the system has low tolerance for deviations from the patterns set by $r_{ij}$. 

We find that our results are reasonably insensitive to the value of $\beta$ as long as it is past the ``critical point'' (around unity). Empirically, we find that the most faithful transmission is possible when $\tau$ is set to be large, and $\beta$ is slowly increased from zero to a value comparable to $\tau$; in the machine learning literature this is known as simulated annealing. Once this process has finished, the final configuration is the practice of the learner; the learner, in turn, can act as a teacher for a new (randomly initialized) learner; the old learner/new teacher intervenes on this new learner in the same fashion.

All that remains is to determine a good candidate for $T$, the set of nodes the teacher intervenes on. We do this in a ``greedy'' fashion. We first find the best facet for a single intervention (\emph{i.e.}, the facet that, if fixed by the teacher, allows the learner to best approximate the desired activity). We then iterate the process: having found this facet, we then find the facet that, when fixed in conjunction with the first, produces the best outcome, and so forth. Over a decade range of network sizes (from ten facets to one hundred), we find that it is necessary to fix between 10\% and 15\% of the facets by teaching.


\section{Acknowledgements}

HM acknowledges the support of an Omidyar Fellowship, and SD acknowledges the support of the Survival and Flourishing Fund and the John Templeton Foundation. We thank Colin Allen, Eddie Lee, Celia Heyes, Mirta Galesic, Paul Hooper, Cailin O'Connor, Paul Smaldino, and Mason Youngblood for helpful conversations.


\bibliographystyle{apacite}
\bibliography{main}

\end{document}